# Prodorshok I: A Bengali Isolated Speech Dataset for Voice-Based Assistive Technologies

A comparative analysis of the effects of data augmentation on
HMM-GMM and DNN classifiers

Mohi Reza, Warida Rashid, Moin Mostakim
Department of Computer Science & Engineering
BRAC University
Dhaka, Bangladesh
mohireza@ieee.org, warida.rashid@gmail.com, mostakim@bracu.ac.bd

*Abstract*— Prodorshok I is a Bengali isolated word dataset tailored to help create speaker-independent, voice-command driven automated speech recognition (ASR) based assistive technologies to help improve human-computer interaction (HCI). This paper presents the results of an objective analysis that was undertaken using a subset of words from Prodorshok I to assess its reliability in ASR systems that utilize Hidden Markov Models (HMM) with Gaussian emissions and Deep Neural Networks (DNN). The results show that simple data augmentation involving a small pitch shift can make surprisingly tangible improvements to accuracy levels in speech recognition.

*Keywords—Automatic Speech Recognition; Bengali; Hidden Markov Model; Gaussian Mixture Model; Deep Neural Network; Human Computer Interaction, Assistive Technology*

## I. Introduction

The impetus for creating Prodorkshok I is twofold. First, it serves to help fill the lack of preprocessed, easy to use Bengali isolated-word datasets. Second, this word set can have useful applications in assistive technologies that utilize voice commands. For example, an application that combines our ASR system with text-to-speech technology can enable people with visual impairment to navigate through digital interfaces with ease.

Contemporary software systems are heavily reliant upon increasingly rich graphical user interfaces. While this has brought drastic improvements in terms of usability for the general population, the same cannot be said for people suffering from disabilities such visual impairment or lack of mobility. Where accessibility features do exist, they are rarely designed with Bengali speaking users in mind. With an estimated 650 000 visually impaired adults in the Bangladesh [1], there is a tangible need for more inclusive alternatives to purely GUI-driven ways to navigate digital interfaces. Prodorshok I can help fill this void.

## II. Related Work

The research landscape for Bengali Speech Recognition is nascent in comparison to the rich history of ASR system development in English. A notable Bengali dataset that is available for free is SHRUTI Bengali *Continuous* ASR Speech Corpus [1]. Das and Mitra [2] used a Hidden Markov Toolkit (HKT) to align its speech data. Mandal et al. [3] used SHRUTI to create a phone recognition (PR) system that used an optimum text selection technique to decipher the smallest discrete unit of sound in uttered speech. In 2010, Mandal, Das and Mitra [4] introduced SHRUTI-II, a SPHINX3 based Bengali ASR System and demonstrated its use in an E-mail based computer application designed to aid visually impaired users. Mohanta and Sharma [5] did a small study on emotion detection in Bengali speech. Their goal was to identify neutrality, anger and sadness in speech using Linear Prediction Cepstral coefficient (LPCC), Mel-frequency Cepstral Coefficient (MFCC), pitch, intensity and formant. Bhowmik and Mandal [16] applied a deep neural network based phonological feature extraction technique on Bengali continuous speech.

## III. Dataset

The dataset consists of recordings of single utterances of 30 Bengali words by 35 native speakers in Dhaka, totaling 1050 voice samples. The word set has been specifically constructed to be used in systems that implement hands-free selection and navigation of digital interfaces. It includes Bengali words for 10 digits (0 to 9), 10 directional words (East West, North and South, up, down, left, right, forward, backward) and 10 positional words (First to tenth).

## IV. Method

### A. Preprocessing

All word samples were put through five stages of speech enhancement. First, stereo channels were merged into mono. Then, static background noise was attenuated using a noise-reduction algorithm based on Fourier analysis. Unique noise profiles for different samples were used for best results. The sound signals were then normalized to have a maximum amplitude of -1.0 dB. The mean amplitude displacement was set to 0.0 for uniformity. Any silence at the beginning or end were truncated. Finally, the audio samples were cloned into two separate datasets, one of which was synthetically augmented by including pitch altered voice samples of existing data.

The effect of these five stages on the audio sample can be seen in fig. 1. The end result is a concise audio sample that is ready to be used to train and test different acoustic models.

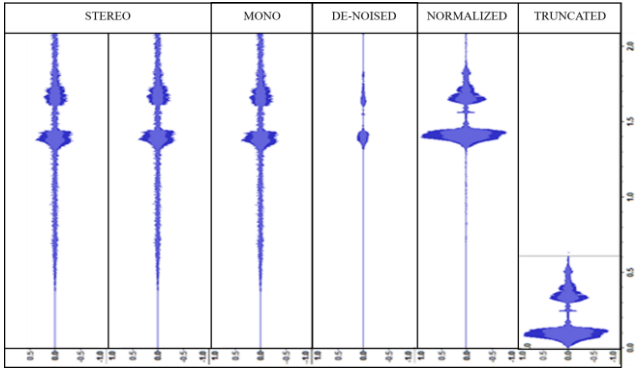

Fig. 1.  Visualizing the Five Stages of Speech Enhancement

*B. Feature Extraction*

After the data was prepared, the next step involved selecting and extracting features from the data. The accuracy and precision of speech recognition systems are highly dependent on the method of feature extraction. For this experiment, the state-of-the-art feature for speech recognition system — MFCC (Mel-Frequency Cepstral Coefficient) was selected. MFCC features closely mimic the way sound is perceived by the human ear. The implementation details [7, 8, 9] are as follows:

- The signal is divided into 25ms long frames.
- For each frame, the Discrete Fourier Transform (DFT) is taken. The DFT is represented by the following equation:

$$X_N[n] = X_N[n + N] \quad (1)$$

Here, $X_N[N]$ is a signal with a period of N. If the DFT is $X_N[n]$, k is the length of the DFT.

$$X_N[k] = \sum_{n=0}^{N-1} x_N[n] e^{-j2\pi nk/N} \quad (2)$$

$$X_N[k] = \frac{1}{N}\sum_{n=0}^{N-1} x_N[n] e^{-j2\pi nk/N} \quad (3)$$

The periodogram estimate of power spectrum is given by:

$$P_N[k] = \frac{1}{N}|x_n[k]|^2 \quad (4)$$

- The Mel-spaced filterbank is applied to the results of the previous step. The filterbank energies are calculated by multiplying each filterbank with the power spectrum and adding the coefficients.
- Then, the logarithm of the filter bank energies is taken.
- Finally, the Discrete Cosine Transform (DCT) of the log filter bank energies are computed to yield the cepstral coefficients. The lower 13 coefficients result in a feature vector for every frame of each signal.

*C. Hidden Markov Model (HMM) with Gaussian Emission*

Hidden Markov Models have been successfully used for time varying sequences such as audio signal processing. The underlying idea behind the Hidden Markov Model is that, it models sequences with discrete states. The way this maps to the problem of speech recognition is, during the feature extraction process, speech signals are transformed into features of discrete time slices or frames. Therefore there are a finite number of frames in a particular word. When a particular sequence of features is given, the model can yield the probability of that sequence being a certain word. Here, the phonemes i.e. the distinct units of sound that can be produced are discrete states and the sequences of MFCCs which represent the uttered word are observations. The probability of observing MFCC sequences given the state is performed using Gaussian Emissions. The details [10] of how it works are stated below:

Hidden Markov Model is a probabilistic model which can produce a sequence of observations X by a sequence of hidden states Z. It is generated by a probabilistic function associated with each state.

An HMM is usually represented by λ where λ = (A, B, Π). It can be defined by the following parameters:

- O = {$o_1, o_2,\ldots\ldots o_m$}. This is an output observation sequence. For speech recognition, this represents the MFCC feature vectors.
- Ω = {1, 2,………,N}. This is a set of states. For speech recognition, it is the phoneme labels.
- A = {$a_{ij}$}. This is the transition probability matrix. It represents the probability associated with transition from state i to state j.
- B = {$b_i(k)$}. It is the output probability, i.e. the probability of emitting a certain observation $o_k$ in the state i.
- Π = Start probability vector.

There are three basic problems for HMM:

- Estimating the optimal sequence of states given the parameters and observed data.
- Calculating the likelihood or probability of a data given the parameters and observed data P (O| λ).
- Adjusting the parameters given the observed data so that P(O| λ) is maximized.

*1) Estimating the Parameters*

The solution to problem 3 is to estimate the model parameters so that P (O| λ) is maximized for the training observations. The optimal Gaussian mixture parameters for a given set of observations can be chosen such that the probability reaches maxima by using the Expectation Maximization (EM) algorithm [11]. It is a gradient based optimization method which is likely to converge at the local maxima.

*2) Estimating the State Sequence*

Estimating the state sequence S given an observation sequence X and the model λ is done using the Viterbi algorithm [12]. It is a formal technique for finding the best state sequence based on dynamic programming method [10].

*D. Deep Neural Network*

Deep Neural Network has proven to be successful in speech recognition and is currently a widely researched area under this field. [13, 14, 15] Artificial neural networks are simplified representation and simulation of the neuronal structure present in brains. Deep neural networks are artificial neural networks where multiple layers of neuron are used. The system learns through observations and the feedback mechanism.

*1) Activation Function*

McCulloch and Pitts proposed the idea of an artificial neuron called the Sigmoid Neuron [16]. The sigmoid function is widely used for feed-forward network with backpropagation because of its non-linearity and simplicity of computation [17]. The function is given by:

$$f(X) = \frac{1}{1+e^{-g(X)}} \qquad (2)$$

In practical applications of using the sigmoid function as an activation function, $W_i$ is a real valued weight. $X_i$ is the input and the weighted input of a node is given by:

$$g(X) = X_1W_1 + X_2W_2 + \cdots + X_iW_i + \cdots + X_nW_n + b \qquad (3)$$

The weight variable is changed depending on how much the relationship between the inputs to the output needs to be strengthened.

*2) Multi-Layered Feed-forward Network*

The Feed-forward network is the type of Artificial Neural Network where connections between the nodes do not form a cycle [18]. A simple three layered feed-forward neural network structure is shown below:

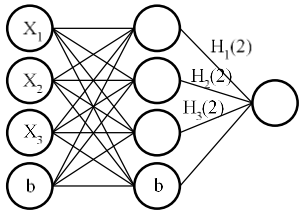

Fig. 2. A three-layered neural network

*3) Backpropagation*

Backpropagation is the process of minimizing differences between actual output and desired output or error based on the training samples with labels. This is used by optimization algorithms for adjusting the weight for each connection between neurons or nodes in different layers based on the accumulated error of a batch in the training data. The error is computed using a cost function which is propagated back through the network. Different optimization algorithms are used for the process.

*4) Optimization Algorithm*

The objective of an optimizer is to get to the minimum point of the error curve for different weights. For our experiment, Adam, a stochastic optimization method was used which combines the advantages of two popular methods AdaGrad and RMSProp. This technique was first introduced in 2014 [13]. It takes the following parameters:

- **Learning rate** – A floating point value. The learning rate. 0.001 was used for the experiment.
- **beta1** = 0.9 – A constant float tensor. The exponential decay rate for the 1st moment estimates.
- **beta2** = 0.999 – Another constant float tensor. The exponential decay rate for the 2nd moment estimates.
- **epsilon** – A small constant for numerical stability.

*5) The Cost Function*

A way of generalizing the optimization process by not overfitting it to the training set is using cost function. For the experiment, softmax cross entropy with logits were used. It measures the probability error in discrete classification where each sample can belong to exactly one class.

## V. RESULTS

The performance of speaker independent dataset without augmentation for both classifiers were quite low. Interestingly, after data-augmentation, the accuracy levels of the HMM-GMM model increased by 6.12% and that of the DNN by 7.65%. The overall performance however was better with the HMM-GMM model. On the speaker dependent system, the performance of the HMM-GMM model is quite high with an accuracy level of 96.67%. The performance score for DNN in the speaker independent system is comparatively lower, at 47.84% with augmentation and 40.19% without. Further experiments show a positive correlation between the number of utterances per word and the accuracy level in both classifiers.

### A. Average Percentage Accuracy Levels

Each sub-category in Table 1 denotes the average accuracy score that were derived from three consecutive runs of a particular classifier. These results are presented visually in Figure 2.

TABLE I. AVERAGE ACCURACY LEVELS

|  |  | Classifiers | |
|---|---|---|---|
|  |  | *HMM-GMM* | *DNN* |
| Speaker Independent | With Augmentation | 56.28 % | 47.84 % |
|  | Without Augmentation | 50.07 % | 40.19 % |
| Speaker Dependent |  | 96.67 % | 43.75 % |

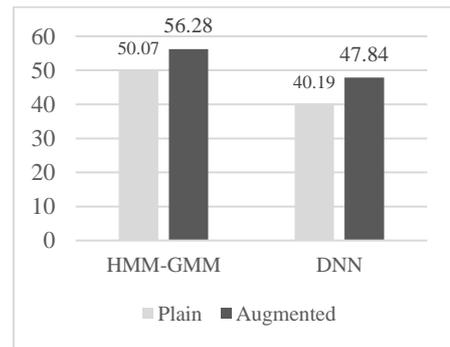

Fig. 3. Effects of Augmentation on Accuracy Levels

### B. Correlation Between Utterences per Word and Accuracy

Table 2 lists how accuracy levels derived from each classifier varied as the number of utterances for each word varied. These results are presented visually in Figure 3.

TABLE II. UTTERANCE PER WORD VERSUS ACCURACY

| Utterances Per Word | | | Classifier | |
|---|---|---|---|---|
| *Total* | *Test* | *Train* | *HMM-GMM* | *DNN* |
| 10 | 3 | 7 | 34.93 % | 21.53 % |
| 15 | 4 | 11 | 35.38 % | 22.89 % |
| 20 | 6 | 14 | 36.97 % | 28.57 % |
| 25 | 7 | 18 | 41.00 % | 31.65 % |
| 30 | 9 | 21 | 50.75 % | 34.63 % |
| 35 | 10 | 25 | 52.51 % | 40.19 % |

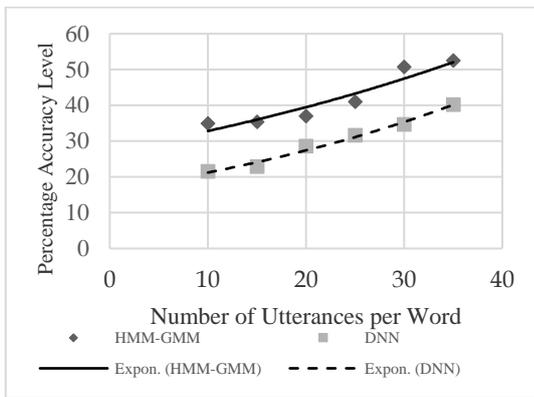

Fig. 4. Plot of Utterance Count and Accuracy Level

## VI. DISCUSSION

Results pertaining to speaker-dependent systems have been very promising, yielding accuracy levels averaging at 96.67% when using the HMM-GMM based classifier. Quite interestingly, the DNN classifier yielded only a negligible improvement over speaker-independent accuracy levels.

The two classifiers we have used are both based on tried and tested acoustic models. Yet, owing to the intrinsic acoustic variability in sound signals spoken by multiple speakers, accuracy levels for speaker-independent systems have been below 60%. The size of the corpus and the sparseness of our training data are what likely affected the performance of the classifiers. Judging by the experimental results summarized in Table 2, there appears to be a clear positive correlation between utterance count and accuracy levels for both classifiers. As such, expanding the current dataset to incorporate higher number of utterances per word will likely solve this issue.

Despite the limitations of working with a sparse dataset, experimental results summarized in Table 1 indicate that tangible improvements can be made by augmenting the data through simple measures such as pitch shifting. This is likely due to the increased variety of speech signals to which the ASR system is exposed to during training.

## VII. FUTURE WORK

The performance of hybrid classifiers such as DNN-HMM is yet to be explored. It would be useful to see if the trend lines depicted in Figure 4. hold as utterance count is increased beyond 35. Empirical analysis of this nature relies upon the availability of more data. Hence, there is a need for further expansion of Prodorshok I to incorporate larger vocabulary and increased variation in speech for every word.

## VIII. CONCLUSION

In this paper, we tested *Prodorshok 1* using two classification algorithms that use HMM-GMM and DNN based acoustic modeling. The results indicate that Prodorshok I in its current form can already be used to design reliable speaker-dependent systems. Furthermore, they show that a simple data augmentation technique relying upon minor pitch shifting can make tangible improvements in speech recognition accuracy. Further expansion to the dataset will likely improve performance levels in speaker-independent systems.

ACKNOWLEDGMENT

We are grateful to Assistant Professor Matin Saad Abdullah and Lecturer Md. Shamsul Kaonain, Department of Computer Science and Engineering at BRAC University for their guidance and engaging conversations that were indispensable to the development of this work.